\documentclass[prl,cha, preprintnumbers ,amsmath,amssymb, showpacs, raggedbottom, floatfix]{revtex4}

\usepackage{bm}
\usepackage[colorlinks=true,linkcolor=blue]{hyperref}
\usepackage{amsmath}
\usepackage{amssymb}
\usepackage{amsthm}
\usepackage{amsfonts}
\usepackage{times}
\usepackage{enumerate}
\usepackage{latexsym}
\usepackage{ifpdf}
\usepackage{graphicx}
\usepackage{color}
\usepackage{makeidx}
\usepackage{ulem}

\expandafter\ifx\csname package@font\endcsname\relax\else
\expandafter\expandafter
\expandafter\usepackage
\expandafter\expandafter
\expandafter{\csname package@font\endcsname}%
\fi
\begin{document}

\title{Electronic Structure of Oxide Interfaces: A Comparative Analysis of GdTiO$_3$/SrTiO$_3$ and LaAlO$_3$/SrTiO$_3$ Interfaces}
\author{Hrishit Banerjee$^1$}
\author{Sumilan Banerjee$^2$}
\author{Mohit Randeria$^3$}
\author{Tanusri Saha-Dasgupta$^{1}$}
\email{t.sahadasgupta@gmail.com}

\address{$^1$Department of Condensed Matter Physics and Material Sciences,
S.N. Bose National Centre for Basic Sciences, JD Block, Sector-III, Salt 
Lake City, Kolkata 700 098, India}
\address{$^2$Department of Condensed Matter Physics, Weizmann Institute of Science, Israel, 7610001}
\address{$^3$Department of Physics, Ohio State University, Columbus, OH 43210, United States} 
\maketitle

\textbf{Emergent phases in the two-dimensional electron gas (2DEG) formed at the interface between 
two insulating oxides have attracted great attention in the past decade. We present ab-initio 
electronic structure calculations for the interface between a Mott insulator GdTiO$_3$ (GTO) and 
a band insulator SrTiO$_3$ (STO) and compare our results with those for the widely studied 
LaAlO$_3$/SrTiO$_3$ (LAO/STO) interface between two band insulators.
Our GTO/STO results are in excellent agreement with experiments, but qualitatively different 
from LAO/STO. We find an interface carrier density of 0.5$e^{-}$/Ti, independent of GTO thickness 
in both superlattice and thin film geometries, in contrast to LAO/STO. The superlattice geometry in 
LAO/STO offers qualitatively the same result as in GTO/STO. On the other hand, for a thin film 
geometry, the interface carrier density builds up only beyond a threshold thickness of LAO.
The positive charge at the vacuum surface that compensates the 2DEG at the interface also exhibits 
distinct behaviors in the two systems. The top GTO layer is found to be insulating due to 
correlation-driven charge disproportionation, while the top LAO layer is metallic within band theory 
and becomes insulating presumably due to surface disorder or surface reconstruction.}


\bigskip

Following the pioneering work by Ohtomo and Hwang~\cite{hwang} on LAO/STO, there has been much effort on understanding the interface (IF) between two different insulating ABO$_3$ perovskites. 
The [001] stacking consists of AO and BO$_2$ layers, which are charge neutral in one of the oxides like the 
(SrO)$^0$ and  (TiO$_2$)$^0$ layers in STO, but have charge +1/-1 in the other oxides, like (LaO)$^{+1}$ and (AlO$_2$)$^{-1}$ 
in LAO. This creates a polar discontinuity at 
the interface of the two types of oxides and a build up of electrostatic potential, which can only be averted by a transfer of charge to the interface. The 2DEG that results from 
this simple {\it polar catastrophe} picture should lead to an interface carrier density of $0.5e^{-}$/Ti, corresponding to $\simeq 3.3 
\times 10^{14} cm^{-2}$, much larger than that achieved in conventional semiconductor hetero-structures. 

The $n$-type interface in LAO/STO heterostructures is the most studied of all oxide interfaces~\cite{manhart,huijben,review1,review2}.
 It exhibits gate-tunable 
superconductivity and in addition shows signatures of local moments and possible ferromagnetism~\cite{brinkman,reyren,bert1,li} coexisting with the superconductivity. 
The density of itinerant carriers at the interface, however, is consistently found to be an order of magnitude smaller~\cite{siemons,basletic,thiel} than
$0.5e^{-}$/Ti, the value expected from the polar catastrophe model. In addition, the interfaces are insulating, rather than being metallic, below a 
certain critical thickness of LAO layers~\cite{thiel}.

A more recent development is the study of the $n$-type interface between the Mott insulator GTO and the band insulator STO grown by molecular beam epitaxy~\cite{moetakef, cain, moetakef1}. Remarkably, the GTO/STO samples give rise to 2DEGs with carrier densities of 
$0.5e^{-}$/Ti, exactly as expected from the ideal polar catastrophe scenario. Furthermore, the GTO/STO interface is found to be conducting 
irrespective of layer thickness of GTO, so that there is no thickness threshold for metallic behavior. In both these respects GTO/STO seems to be qualitatively different from LAO/STO. The 2DEG at the GTO/STO interface also shows many other interesting 
properties~\cite{moetakef,cain,moetakef1}, including possible transport signatures for magnetism~\cite{moetakef2} and 
strong correlations~\cite{moetakef3}.

Motivated by these developments, we present here a detailed electronic structure study of the GTO/STO interface and contrast our results with those obtained for LAO/STO. While the LAO/STO interface has been thoroughly studied by electronic 
structure calculations~\cite{satpathy,lao-elec,pentcheva,chen},
 much less is known about GTO/STO. The specific problem of single SrO layer in a GTO matrix in the 
 superlattice geometry has been studied~\cite{chen2013,walle,lechermann} by a variety of techniques, first-principles, model Hamiltonian as well as combined density functional theory (DFT) and dynamical mean field theory. There have been some suggestions~\cite{janotti} about the origin of the differences between the LAO/STO and GTO/STO systems, but to the best of our knowledge, no first principles electronic structure study exists which compares the LAO/STO and GTO/STO interfaces on same footing in 
different heterostructure geometries. Gaining insight into GTO/STO and into the differences and similarities with LAO/STO is very important for the advancement in the field of oxide interfaces. 

We conclude the introduction by summarizing our main results for $n$-type interfaces in GTO/STO and LAO/STO 
in two different geometries: (a) a superlattice and (b) a thin film on an STO substrate with vacuum on top.
(1) Both GTO/STO and LAO/STO show essentially similar behavior in the superlattice geometry, despite differences in details of the orbital character 
of the carriers due to different structural distortions. The key result is that both GTO/STO and LAO/STO superlattices have an interfacial charge 
density of 0.5 $e^{-}$/Ti, and there is no critical thickness of GTO or LAO for conductivity in the superlattice geometry. (2) The behavior of the two 
systems is very different in the thin film-substrate geometry. We find that GTO/STO conducts even for GTO thickness of 1 unit cell, the minimum thickness possible, while LAO/STO conducts only beyond a critical thickness of 5 unit cells of LAO. (3) In the thin film-substrate geometry, GTO/STO has an interface carrier density of 0.5 $e^{-}$/Ti independent of GTO thickness. In contrast, the carrier density in LAO/STO is about a factor of four smaller just 
beyond the threshold for conductivity, and rises slowly with increasing LAO thickness. (4) We find that, in the thin film geometry,  the surface layers 
facing the vacuum have very different electronic structures in LAO/STO and GTO/STO.
The holes on the top GTO layer are found to be localized due a correlation-driven charge disproportionation on Ti,
while the holes on top LAO layer are metallic within band theory.

\section{Results}

\subsection{Heterostructure Geometries}

A study of both experimental and theoretical 
literature~\cite{hwang,manhart,huijben,moetakef,cain,moetakef1,lit,zhang01,zhang02} shows that the oxide 
interfaces have been investigated in two different geometries, (i) superlattice geometry with periodic  repetition of alternating layers of STO and, LAO or GTO, and (ii) thin film of LAO or GTO grown on a STO(001) substrate. Most experimental studies on LAO/STO are carried out in
a thin film-substrate geometry, while both geometries have been investigated in GTO/STO experiments.
Since we would like to have a comparative study of the two systems with an aim to arrive at a common
understanding, in the present study we consider both the geometries, as shown in Fig.~1. 
Following the experimental literature on 
GTO/STO~\cite{moetakef, cain, moetakef1}, we consider only $n$-type IFs formed between GdO 
layer from GTO and TiO$_2$ layer from STO in GTO/STO, and between LaO and TiO$_2$ in LAO/STO. 
In the superlattice geometry, we consider two symmetric $n$-type interfaces in the cell 
which result in non-stoichiometric supercells with an additional TiO$_2$ layer in STO and an additional GdO layer in GTO 
(LaO layer in LAO). 
This results in superlattices with the formula (LAO)$_{p.5}$/(STO)$_{q.5}$ or (GTO)$_{p.5}$/(STO)$_{q.5}$.  Calculations are carried out 
for $p$ = 1, 2, 3, 4 and $q$ = 4 and 9.

The thin film-substrate geometry, shown in the bottom panel of Fig.~1, creates a single $n$-type interface and a surface of 
TiO$_2$ (in case GTO/STO) or AlO$_2$ (in case of LAO/STO) facing the vacuum. The general formula of the thin film-substrate systems is 
(LAO)$_{p}$/(STO)$_{q}$ or (GTO)$_{p}$/(STO)$_{q}$. Calculations are carried out for $p$ = 1, 2, 3, 4, 5, 6 and $q$ = 9. 

The in-plane dimension of the simulation cell is expanded by $\sqrt{2} \times \sqrt{2}$ creating two Ti or Al atoms in the BO$_2$ layers
of the unit cell to take into account the GdFeO$_3$-type orthorhombic distortion characterized by tilt and rotation of the TiO$_6$/AlO$_6$
octahedra. This becomes specially important for the GTO/STO system, as we will see in the following.

\subsection{Structure}

We first start with discussion of the structural properties of the studied heterostructures. As already mentioned, the presence of 
GdFeO$_3$-type orthorhombic distortion is an important structural aspect of GTO. This distortion in bulk GTO makes the 
structural properties of the optimized GTO/STO systems rather different compared to LAO/STO. Structural distortions
observed include the tilt and rotation of the metal (M) - oxygen(O) octahedra as well as the 
compression or elongation of the individual MO$_6$ octahedra. Fig. 2 shows the plots of the deviation of 
M-O-M bond angle from 180$^\circ$, as well as the difference between out-of-plane and in-plane M-O bond lengths. The former 
quantifies the tilt/rotation of MO$_6$ octahedra, while the latter quantifies the compression (for negative sign) or 
elongation (for positive sign) of MO$_6$ octahedra. 
The top panels of the figure show the result for the superlattice
geometry while the bottom panels are for the thin film-substrate geometry. The qualitative behavior is similar between
the two geometries. 

For the GTO/STO systems, the deviation of the  Ti-O-Ti bond angle from 180$^\circ$ is as 
high as 30$^\circ$ or so in the GTO side. This decreases systematically and reaches a value of about 5$^\circ -$ 10$^\circ$ 
within the interior of STO block. For thin film-substrate geometry the tilt/rotation attains a constant value within
the interior of STO, which is found to be substantial for rotation. 
The TiO$_2$ layer at the IF faces the SrO layer on one side and GdO on the other. This makes the tilt angle
in the $+c$ and $-c$ directions with respect to the TiO$_6$ octahedra to be different
($c$-axis is the [001] direction that is perpendicular to the IF).
This effect is found to percolate to other TiO$_2$ layers as well, specially in the case of GTO/STO.
At the interfaces, the highly 
asymmetric out-of-plane tilt angles vary between about 15$^\circ$ and about 25-30$^\circ$. The in-plane and out-of-plane Ti-O 
bondlengths become unequal in GTO layers, with maximum difference of 0.1 - 0.2 $\AA$, indicating 
distortion of the TiO$_6$ octahedra. This distortion becomes smaller at IF and inside the STO block it attains a 
value of $\approx$ 0.1 $\AA$ or smaller. 
The MO$_6$ octahedra are compressed in GTO layers, and are elongated in 
STO block. For the thin film-substrate geometry, the distortion of TiO$_6$ attains more or less a small constant value
inside the interior of STO block.

In comparison, in LAO/STO, the deviation of M-O-M bond angle from 180$^\circ$ occurs only for rotation, which is much
smaller in magnitude compared to GTO/STO. The rotation angles are only significant at the IFs or close to them 
with values of about 5-9$^\circ$. The tilt angles are found to be zero. Like in GTO/STO, the 
metal-oxygen octahedra are compressed in LAO side and elongated in STO side. 
The structural differences between GTO/STO and LAO/STO, specially in terms of tilt and rotation of MO$_6$ octahedra, has
important bearing on the orbital character of the conducting electrons at different layers, as will be elaborated
in the following.

\subsection{Electronic Structure}

We analyze the electronic structure of the optimized GTO/STO and LAO/STO heterostructures in both superlattice and 
thin film-substrate geometry in terms of density of states, charge and orbital populations.
Fig. 3 shows the layer-wise density of states (DOS) projected to Ti $xy$, $xz$ and $yz$ states in 
(GTO)$_{1.5}$/(STO)$_{4.5}$ and (LAO)$_{1.5}$/(STO)$_{4.5}$ superlattices. Qualitatively similar results are 
obtained for (GTO)$_{1.5}$ or (LAO)$_{1.5}$/(STO)$_{9.5}$ superlattices, proving that the
physics in the superlattice geometry is independent of the STO thickness. First, we find both GTO/STO and 
LAO/STO superlattices are metallic with non-zero density of states at the Fermi level (E$_F$). Since the layer thickness of 
GTO or LAO of 1.5 layers is the minimum possible within the superlattice geometry, we conclude that for both LAO/STO and 
GTO/STO superlattices, there is no minimum thickness for conductivity. As expected, calculations with larger 
thickness of GTO and LAO (checked with thicknesses of 2.5, 3.5 and 4.5 unit cells), are also found to be metallic. 

We find that in both superlattices, the conducting electronic charge is not strictly confined to the IF, and 
spreads out into several layers of the STO block. This is in agreement with experimental findings~\cite{prl_2010},
and previous theoretical studies~\cite{satpathy} of LAO/STO.
We note, however, that the nature of the conduction electrons is different in the two superlattices, as seen from
the DOS in Fig. 3. We find that, while the carriers at the IF are predominantly of $xy$ character
for both GTO/STO and LAO/STO,  the situation is different in two systems in the other TiO$_2$ layers in STO. 
The orbital characters of the carriers in these layers are rather mixed in case of GTO/STO, and mostly $xz/yz$ in case of LAO/STO. 
This difference stems from the structural differences between GTO/STO and LAO/STO. 

To obtain the layer-wise contribution to the conduction electron, we integrate the layer projected DOS from  0.5 eV below E$_F$ to E$_F$.
For GTO/STO system this corresponds to integrating from the upper edge of the lower Hubbard band of Ti $d$ in insulating GTO layer to E$_F$.
The electron from Ti$^{3+}$ ion in GTO layer in its $d^{1}$ charge state occupies the localized lower Hubbard band and 
does not contribute to conduction. 
Table I shows the layer-wise contribution to the conduction electron for 1.5/4.5 as well as 1.5/9.5 superlattices. We find the total 
conduction charge in the LAO/STO as well as in GTO/STO superlattices to be 1 $e^-$ irrespective of the thickness of 
STO layers. This is fully consistent with the presence of two symmetric interfaces each with a carrier density of $0.5e^{-}$/Ti.

A similar analysis in the thin film-substrate geometry shows dramatically different behavior. The left
and middle panels in Fig. 4 show the DOS in different TiO$_2$ layers of LAO/STO
in thin film-substrate geometry with two different thicknesses of 
LAO layers, 1 and 5 unit cells, respectively. We find that the IF in 1 unit cell of LAO on STO is insulating with a large gap 
between the valence and conduction states, and no states at E$_F$. On the other hand, 5 unit cell of LAO on STO 
is barely metallic, setting a critical LAO thickness of 5 unit cell for the conductivity. This behavior is significantly different
from that of LAO/STO in superlattice geometry for which IF's are found to be conducting for any thickness of LAO.
This difference in conduction properties of  LAO/STO, depending on the system geometry has been pointed out
previously in literature~\cite{pentcheva,chen}. 

A markedly different picture is obtained for GTO/STO system. The right most
panel of Fig. 4, shows the plot of density of states of GTO/STO in thin film-substrate geometry with 1 unit cell 
thickness of GTO. We find the solution to be metallic even in the limit of 1 unit cell thickness of GTO. This is in sharp 
contrast to LAO/STO case, but in excellent agreement with experimental reports on GTO/STO~\cite{moetakef}. 
The calculation of total conduction charge by integrating the layer wise density of states from -0.5 eV below E$_F$ to E$_F$ gives a 
charge of $0.5 e^-$ (see Table I) for GTO/STO in thin film-substrate geometry with 1 unit cell 
thickness of GTO. This is in complete accordance with a single $n$-type interface in the unit cell, and 
the expectation from polar catastrophe model. In contrast, the total conduction charge for the 
LAO/STO in thin film-substrate geometry with 5 unit cell thickness of LAO, which is at the critical thickness 
of metallicity, is found to be about  0.14 $e^{-}$/Ti, almost a factor of 4 smaller than 0.5 $e^{-}$/Ti 
expected from the `ideal' polar catastrophe scenario. 
Increasing the LAO thickness beyond 5 u.c., the carrier concentration is found to slowly
increase, for example for (LAO)$_6$/(STO)$_9$ the conduction charge is found to be 0.18 $e^{-}$. 
This is expected to reach the asymptotic value of 0.5 $e^{-}$ for very large thickness of LAO, as discussed below.

The analysis of orbital population of the conduction electron shows the carriers 
at the IF are predominantly of $xy$ character, while that within the STO block to be predominantly of 
$yz/xz$ character in LAO/STO with LAO thickness beyond the critical thickness of conductivity. For GTO/STO,
the IF is of significant $xy$ character, the subsequent layers being of mixed character which converts to 
predominant $yz/xz$ character in the interior of STO block. We thus find that the
orbital characters in LAO/STO and GTO/STO are quite similar in the superlattice and thin film-substrate geometries,
as might be expected given the similarities in the structural distortions in the two geometries.

Another pertinent issue in the context of thin film-substrate geometry is the fate of the surface layer facing the vacuum, 
which is AlO$_2$ in LAO, or TiO$_2$ in GTO. By simple charge balance, the uppermost surface layer should be 
hole doped to compensate for the electrons at the n-type interface. For instance, the surface layer in GTO should have
Ti in a $d^{0.5}$ state, rather than the $d^{1}$ state in bulk GTO. This simple picture, of course, does not
take into account the disordering effects like oxygen vacancies
and cation disorder, or the effect of surface reconstruction, which are reported to be important in the context of 
LAO/STO\cite{adv,zunger}.
Interestingly, our DFT calculation which allows for possible structural reconstruction only within the scope of $\sqrt{2} \times \sqrt{2}$ cell, shows
the topmost AlO$_2$ surface in LAO to be metallic, while the topmost TiO$_2$ surface in GTO to be insulating.
This is seen in the plot of charge density contributed by a narrow energy window around E$_F$, for (LAO)$_5$/(STO)$_9$ and for 
(GTO)$_1$/(STO)$_9$ (cf Fig. 5). 
While the surface reconstruction in reality can be complex, which undoubtedly needs further exploration both 
from experimental and theoretical side, the stabilization of the insulating solution at the topmost TiO$_2$ surface layer of
GTO doped with 0.5 hole is interesting. We find this to be triggered by charge disproportionation between two Ti atoms at the
top layer. With the choice of $U$ = 7 eV on Ti atoms (see Methods for further discussion), 
this charge disproportionation becomes complete leading to 
insulating solution with nominal charge of $d^{1}$ on one Ti atom and $d^{0}$ on the other, maintaining an average charge of $d^{0.5}$ per 
Ti at the top layer. This is evident from the charge density plot focused on the energy window around the occupied lower Hubbard band (LHB) 
of Ti $d$ states (see inset in Fig. 5), which shows significantly large charge on one set of Ti atoms and a significantly smaller on the other. 
We find this disproportionation to be triggered by the strong correlation effect which together with 
small differences in local environment of two Ti atoms, make the charges on two Ti's significantly different. 

We show the DFT layer-resolved density of states, projected onto O $p$, Ti $d$ and Gd $d$ states, for 
(GTO)$_{2}$/(STO)$_9$ in the thin film-substrate geometry in the right panel of Fig.~6. 
To clearly show the charge disproportionation in this plot, we plot the DOS contributions for the two 
inequivalent Ti atoms at the surface using two different colors in Fig.~6.
The disproportionation weakens as we move away from the vacuum surface towards the interface and is absent at the IF. 
We note that the
charge disproportionation found at the surface does not lead to an additional translational symmetry breaking,
associated with the opening of the charge gap, given that a $\sqrt{2} \times \sqrt{2}$ structural distortion is already present in bulk GTO.

\subsection{Electronic Reconstruction}

Our DFT results described above lead to the following conclusions.

{(i)} Both GTO/STO and LAO/STO show essentially similar behavior in the superlattice geometry, despite differences in 
details of the orbital character of the carriers due to differences in the structural distortions. The central result is that 
both GTO/STO and LAO/STO superlattices have the full interfacial charge density of
$0.5e^{-}$/Ti and there is no critical thickness of GTO or LAO for conductivity.

(ii) The behavior of the two systems differs qualitatively in the thin film-substrate geometry. We find that GTO/STO conducts 
with an interfacial carrier density of $0.5e^{-}$/Ti, independent of GTO thickness. In contrast,  
LAO/STO conducts only beyond a critical thickness of 5 unit cells of LAO, and the interfacial carrier density
is still almost a factor of four less than $0.5e^{-}$/Ti just beyond threshold.

We now address the question: What are the driving mechanisms behind the similarity and differences 
between the two geometries in the two systems? 

In the superlattice geometry, we need an extra GdO or LaO layer with a charge of +1, 
as emphasized by Chen {\it et al.}~\cite{chen}.
This is then directly responsible for doping the two (symmetric) $n$-type interfaces with $0.5 e^{-}$/Ti each.
In effect, there is no polar catastrophe (potential divergence) that is averted in this geometry;
rather one simply obtains a fixed interface carrier concentration, independent of the thickness of GTO or LAO,
as demanded by charge neutrality.

However, in the film-substrate geometry with a single interface, we find that band alignment, band bending and 
electronic reconstruction play a crucial role. As a result, there are striking differences between
LAO/STO and GTO/STO. To understand these differences, it is useful to look at the DFT results
for the density of states (DOS) in the thin film geometry.

The left panel in Fig.~6 shows the density of states plotted over a wide energy window, 
projected onto O $p$, Al $p$, Ti $d$ and La $d$ states for (LAO)$_{5}$/(STO)$_9$ in the thin film-substrate geometry. 
This provides the picture after electronic reconstruction as given by DFT. 
We see that the upper edge of the oxygen valence band (VB) in LAO bends progressively towards E$_F$,
moving from the interface to the surface. We see that the surface AlO$_2$ layer next to vacuum is hole doped
into a metallic state with electrons transferred from the surface to the conduction band (CB) of Ti $d$-states at the IF.
This can be seen schematically in the lower left panel of Fig.~7.

The band alignment before electronic reconstruction can be derived from the bulk band structure of LAO and STO shown schematically in 
the upper left panel in Fig.~7. The experimentally measured bulk band gap of STO is 3.3 eV, while that of LAO is 5.6 eV.
In LAO/STO, the valence band offset is small, and the VB maxima of LAO and STO, 
which is the upper edge of filled O $p$ bands, are almost aligned.  
For the electronic reconstruction, therefore a large band bending with VB maxima of LAO at the surface aligning with 
conduction band minima of STO is needed, as observed in DFT results presented in Fig.~6 and
schematically in left lower panel of Fig.~7. 

The necessary band bending in LAO/STO is estimated to be about the same as band gap $\Delta=3.3$ eV of STO. 
One can make a simple estimate~\cite{pentcheva,millis}
of the critical thickness $N_c$ of LAO layers for charge transfer from the surface to the interface as follows.
The potential difference between the surface and interface is $eE_\mathrm{pol}N_c a$, due to the polar field $E_\mathrm{pol}=2\pi e/\epsilon a^2$
(where $\epsilon$ is the dielectric constant of LAO arising from both electronic and ionic screening) should be equated to the gap $\Delta$. 
Using the bulk dielectric constant of LAO, $\epsilon\simeq 24$, and the in-plane lattice constant of $a=3.9~\AA$, the critical thickness could be estimated to be $\sim 4$ layers, consistent with experiment and the DFT result. For thickness $N>N_c$, when an amount of charge $q$ is transferred to the interface the potential difference $ (4\pi e/\epsilon a)(0.5e-q)N$ stays pinned at $\Delta$. This leads to an estimate of $q(N)=0.5 e(1-N_c/N)$, that approaches the asymptotic value of 0.5$e$ for $N\to \infty$. 
For 5 LAO layers, one obtains $q\simeq 0.10$, roughly consistent with the DFT result of Table I.

The electronic reconstruction scenario is quite different for GTO/STO. Bulk GTO is a Mott insulator
and the relevant valence band is the Ti $d$ lower Hubbard band (LHB), which lies far above the O $p$ derived VB of the STO, 
as shown schematically in the right upper panel in Fig.~7. The Ti $d$ LHB lies only $\approx 0.5$ eV below the conduction band of STO.
Electrostatic considerations, similar to the ones discussed above for LAO/STO, with the bulk dielectric constant of 30 for GTO imply 
that a minimum thickness of 1 u.c. of GTO is sufficient to allow for required band bending and charge transfer. 
This would suggest the charge is transferred to the interface independent of the thickness of the GTO layers, 
consistent with our DFT results and also with the experimental observations~\cite{moetakef,cain}.
Apart from the usual upward bending of the GTO VB approaching the surface, the Ti $d$ LHB in GTO also bends upward near the interface to connect with the Ti $d$ derived conduction band in the STO side, as seen in Fig.~6 (right panel). 

Finally, we stress again the fate of the holes on the top surface layer next to vacuum, that must exist to counterbalance the electrons at the 
interface. Unlike LAO, which is a band insulator, GTO is a correlation driven Mott insulator in the bulk.  As described above, we find a rather strong correlation-induced charge disproportionation on the topmost TiO$_2$ layer of GTO at the surface.
We thus find the LHB of Ti $d$ states at the top layer of GTO, which would naively have been partially filled
(average filling of $d^{0.5}$) and metallic, splits into occupied and unoccupied bands due to opening of a charge gap
due to the charge disproportionation. This is clearly seen in the DFT result of Fig.~6 (right panel) and shown schematically in the lower right panel
of Fig.~7. Our theoretical observation of
charge disproportionation in top TiO$_2$ layer in GTO/STO in thin film-substrate geometry should be explored experimentally.

\section{Discussion and Outlook}
In this work we have carried out a detailed first-principle DFT study of LAO/STO and GTO/STO heterostructures, focussing on their essential similarites and differences in two experimentally well-studied geometries, namely the superlattice and thin film-substrate geometry. While the two systems behave quite similarly in the superlattice set up, e.g.~in terms of the total 0.5 $e^-$/Ti transfer of charges to the interface, very diffrent pictures emerge in case of thin film-substrate geometry due to the differences in electronic reconstructions in the two systems, even though, in both the cases, the reconstructions are driven by the same underlying electrostatic mechanism, namely the need to avert the `polar catastrophe'. We find a full 0.5 $e^-$/Ti conducting charge at the interface even for 1 u.c.~thick GTO on STO substrate, consistent with experiments.\cite{moetakef,cain} On the other hand, in case of LAO on STO, the transferred charge only increase gradually with thickness from a small value of $\sim 0.14$ e$^-$/Ti above a critical thickness of about 4.     

Additionally, in the thin film-substrate case, we find the fate of the surface layers, which host the neutralizing charges for the interface carriers, to be quite distinct. The electronic states derived from O p  orbitals at the surface LAO layer turn out 
to be metallic within DFT but are experimentally\cite{adv} found to be localized,
possibly due to disorder or surface imperfections like oxygen vacancies.\cite{zunger}
On the other hand, the Ti $d$ states at the top most layer of GTO/STO correspond to a doped Mott insulating layer of GTO and stays insulating by opening a charge gap via an interesting correlation driven charge disproportionation. Such a correlation induced phenomenon could be a robust feature, at least within a short length scale, even in the presence of surface disorder and, in principle, could be probed experimentally. This also pertains to the experimental verification of the polar catstrophe mechanism by detecting the counter charges at the surface of the interace system. The presence of counducting charges at the interface by itself does not ambiguously establish the polar catastrophe mechanism for the polar oxide interfaces as there could be other possible sources of interfacial charge carriers, e.g.~ oxygen vacancies~\cite{hwang}.             

 A few interesting future directions for theoretical study would be to investigate in detail the sub-band structures, that is relevant for quantum oscillation measurements~\cite{moetakef1}, and spin-orbit coupling in the GTO/STO interface. Our structural analysis indicates substantially larger polar distortions of the Ti-O-Ti bonds at the GTO/STO interface compared to that in LAO/STO. This could potentially lead to much larger Rashba spin-orbit coupling for GTO/STO heterostructure. The magnetic property of the interface 2DEG in GTO/STO should also be investigated by DFT. GTO/STO shows signature of ferromagnetism that could be an intrinsic correlation driven phenomenon, independent of the 
 proximity to magnetic GTO layer~\cite{moetakef2}.   
 
\section{Methods}

Our first-principles calculations are based on plane wave basis as implemented in the
Vienna Ab-initio Simulation Package (VASP)~\cite{kresse, kresse01} with projector-augmented
wave (PAW) potential~\cite{blochl}. The exchange-correlation functional is chosen to be that given
by generalized gradient approximation (GGA)~\cite{pbe}. Correlation effects beyond GGA are taken
into account through an on-site Hubbard $U$ correction in form of GGA+$U$~\cite{dudarev}.
The use of GGA+$U$ turn out to be crucial for the correct description of the Mott insulating behavior
of GTO. The GGA+$U$ implementation in the plane wave code of VASP with PAW potential needed a
$U$ value of 7 eV for the bulk GTO to be insulating. The Hund«s coupling parameters $J_H$ is chosen
be 1 eV. We have thus consistently used $U$ = 7 eV and $J_H$ =1 eV throughout our calculations. We
found a smaller $U$ value ($\approx$ 4 eV) within the linear muffin tin orbital (LMTO) basis calculation to be
sufficient to drive bulk GTO insulating, a value more consistent with spectroscopic considerations~\cite{imada}.
The fact that different $U$ values are needed in different basis set implementations of DFT has been
appreciated in the literature (see, e.g., ref.~\cite{valenti}).

We found in our LMTO calculations of bulk GTO that taking into account the effect of $U$ on both Ti $d$ and Gd $f$ states 
resulted into a ferrimagnetic ground state with antiparallel alignment of the Ti and Gd spins, consistent with
experiment. However, since we are not interested in the magnetism of Gd spins in the present study, 
we used a plane-wave calculation for the results reported here, with Gd $f$ electrons considered to be part of
the core orbitals.  The in-plane lattice constants of the simulation cells are fixed at the experimental lattice
constant of STO with a value of 3.91 $\AA$, while the out-of-plane lattice constant is allowed to relax. The
consideration of tilt and rotation of metal-oxygen octahedra becomes rather important for GTO with significant
orthorhombic distortion. In order to take into account of that, the in-plane dimension of the simulation cell is
expanded by $\sqrt{2} \times \sqrt{2}$. Internal positions of the atoms are allowed to relax until the forces
become less than 0.01 eV/$\AA$.  


\section{Acknowledgments}
HB and TSD would like to thank Thematic Unit of Excellence on Computational Materials Science, funded by Department of Science and Technology for computational facility used in the present study. M.R. would like to acknowledge the support of the
CEM, an NSF MRSEC, under grant DMR-1420451, which also made possible the collaboration with TSD.

\newpage

\begin{table}
\caption{The layer-wise contribution of the conduction electron in GTO/STO and LAO/STO in superlattice
and thin film-substrate geometries. In case of superlattices, results for both (LAO or GTO)$_{1.5}$/(STO)$_{4.5}$
and (LAO or GTO)$_{1.5}$/(STO)$_{9.5}$ are shown. For thin film-substrate geometry, we show the results for LAO (5 u.c.) and GTO (1 u.c.),
the minimum thicknesses in each case at which the 2DEG is formed.}

\begin{tabular}{|c|c|c|c|c|c|}
\hline 
\multicolumn{3}{|c|}{\bf{LAO/STO}} &\multicolumn{3}{|c|}{\bf{GTO/STO}}  \\ 
\hline 
\multicolumn{2}{|c|}{{Superlattice}}& Thin film-substrate &\multicolumn{2}{|c|}{{Superlattice}} & Thin film-substrate  \\ 
\hline 
(LAO)$_{1.5}$/(STO)$_{4.5}$ & (LAO)$_{1.5}$/(STO)$_{9.5}$ &  (LAO)$_5$/(STO)$_9$ &(GTO)$_{1.5}$/(STO)$_{4.5}$ & (GTO)$_{1.5}$/(STO)$_{9.5}$ & (GTO)$_1$/(STO)$_9$\\ 
\hline
Layer: Charge & Layer: Charge & Layer: Charge & Layer: Charge & Layer: Charge & Layer: Charge \\
\hline 
 2 : 0.185 & 2 : 0.200& 6 : 0.062   &  2 : 0.181 & 2 : 0.163 & 2 : 0.185 \\
(1$^{st}$ IF) & (1$^{st}$ IF)& (IF) & (1$^{st}$) IF & (1$^{st}$ )IF & (IF) \\
\hline 
 3  : 0.184 & 3 : 0.104 & 7 : 0.050 & 3 : 0.203 & 3 : 0.101 & 3 : 0.108 \\ 
\hline 
4 : 0.260 & 4  :  0.070 & 8 : 0.006 & 4 : 0.231 & 4 :  0.058 & 4 : 0.034\\ 
\hline 
5 : 0.184 & 5 : 0.066 & 9 : 0.007 & 5 : 0.203 & 5 : 0.082 & 5 : 0.045 \\ 
\hline 
6 : 0.185 & 6 : 0.060 & 10 : 0.005 & 6 : 0.181 & 6 : 0.096 & 6 : 0.049\\
(2$^{nd}$ IF) & & & (2$^{nd}$ IF)& &\\
\hline 
 & 7 : 0.060 & 11 : 0.004 & & 7 : 0.096 & 7 : 0.044 \\ 
\hline 
 & 8 : 0.066 & 12 : 0.002 & & 8 : 0.082 & 8 : 0.029 \\ 
\hline 
 & 9 : 0.070 & 13 : 0.001 & & 9 : 0.058 & 9 : 0.006\\ 
\hline 
 & 10 : 0.104 & 14 : 0.000  & & 10 : 0.101 & 10 : 0.000\\
 & &  & & & \\
\hline
 & 11 : 0.200 & &  & 11 : 0.163 & \\
 & (2$^{nd}$ IF) & & &  (2$^{nd}$ IF)&\\
\hline 
Total Charge: 0.999 & Total Charge: 1.000 & Total Charge: 0.137  & Total Charge: 0.999  & Total Charge: 1.000 & Total Charge: 0.500 \\ 
\hline 
\end{tabular} 
\end{table}

\begin{figure}
\includegraphics[width=0.7\textwidth]{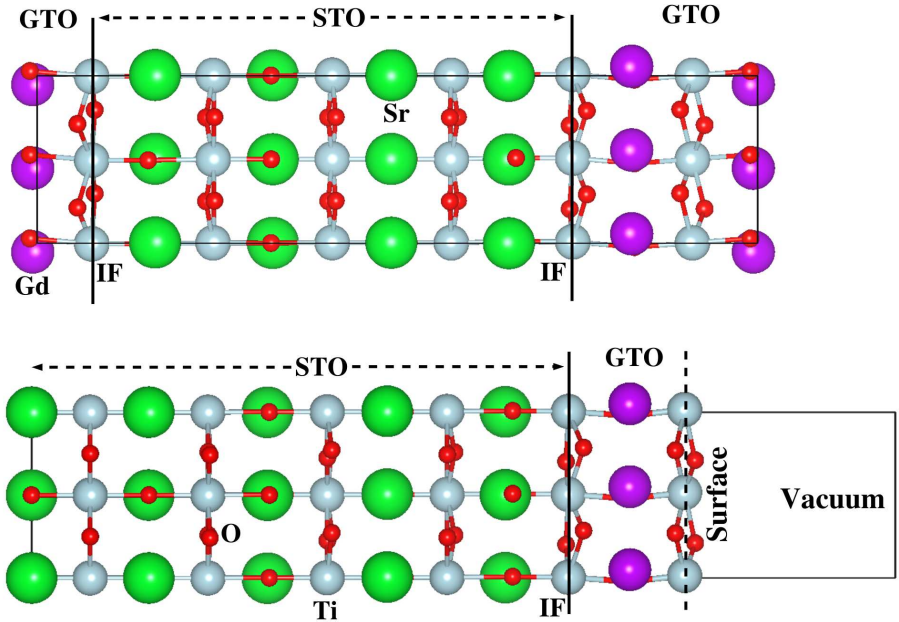}
\caption{The two geometries used in the present study: superlattice (upper panel) and thin film-substrate (lower panel). 
We show the representative cases of (GTO)$_{1.5}$/(STO)$_{4.5}$ superlattice and (GTO)$_1$/(STO)$_5$ thin film-substrate geometry.
The various atoms are: Gd (purple), Ti (grey), O (red), and Sr (green).
The interfaces (IFs) formed between GdO 
from GTO and TiO$_2$ from STO are marked.}
\end{figure}

\begin{figure}
\centering
\includegraphics[width=0.65\textwidth]{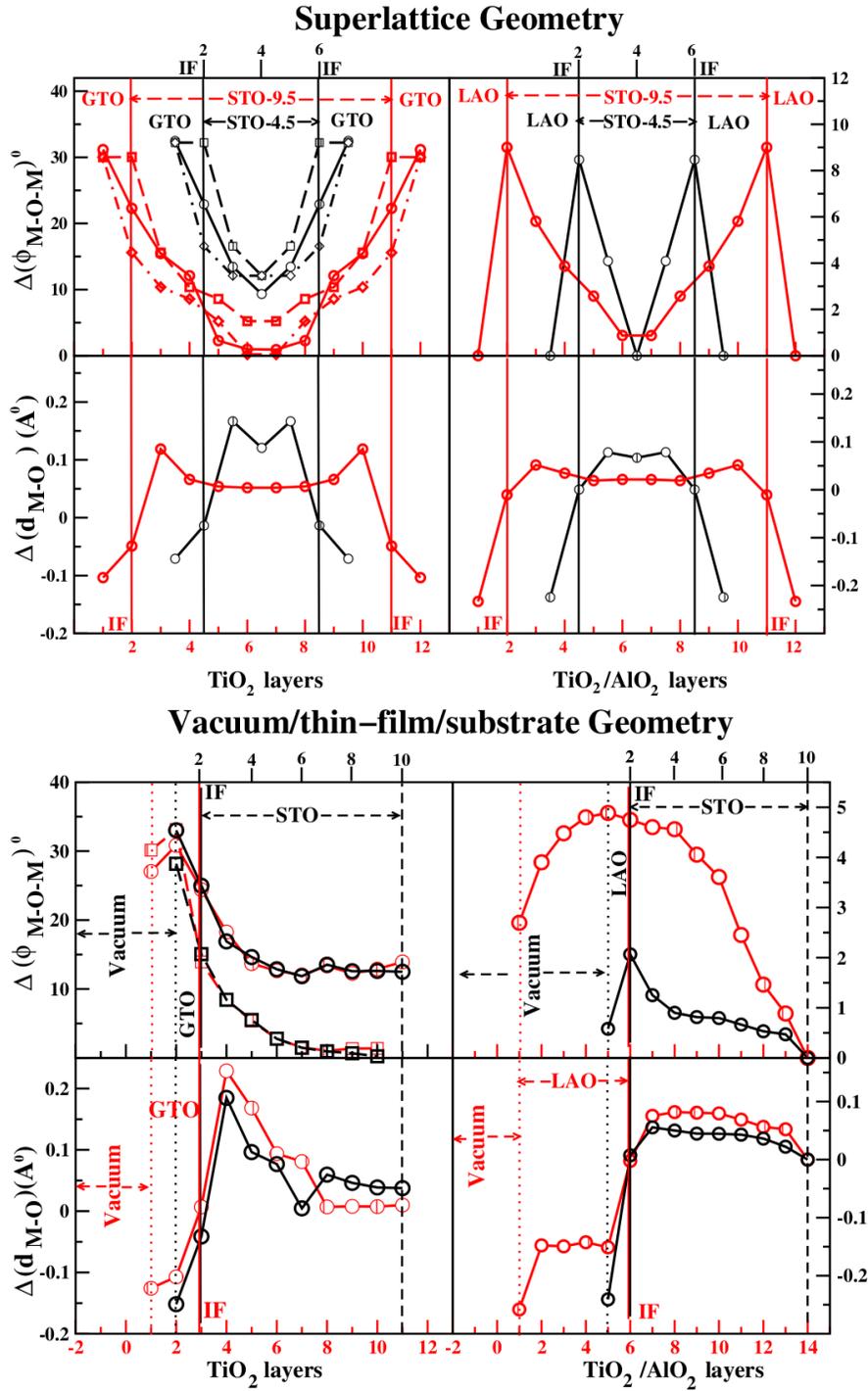}
\caption{The deviation of M-O-M angles from 180$^\circ$, $\Delta$ ($\phi_{M-O-M}$) and the difference of M-O lengths in the 
out-of-plane and in-plane directions, $\Delta$ (d$_{M-O}$).
On the right we show results for LAO/STO and on the left for GTO/STO. 
For superlattices (upper panels), we show results for 
(LAO or GTO)$_{4.5}$/(STO)$_{1.5}$ (black symbols) and (LAO or GTO)$_{9.5}$/(STO)$_{1.5}$ (red symbols). 
For the thin film-substrate geometry (lower panels), 
results are shown for (GTO)$_{1}$/(STO)$_{9}$ (black symbols), (GTO)$_{2}$/(STO)$_{9}$ (red symbols), and 
(LAO)$_{1}$/(STO)$_{9}$ (black symbols), (LAO)$_{5}$/(STO)$_{9}$ (red symbols). 
The x-axis labels for the red curves are marked in red at the bottom of the figures, while
those for the black curves are marked in black at the top.
For GTO/STO, we show the rotations (circles) and the asymmetric tilts in +$c$ (squares) and -$c$ (diamonds) directions; see text.
For LAO/STO only the rotation angles are shown as circles, the tilt angles being zero.  
}
\end{figure}  

\begin{figure}
\includegraphics[width=0.7\columnwidth]{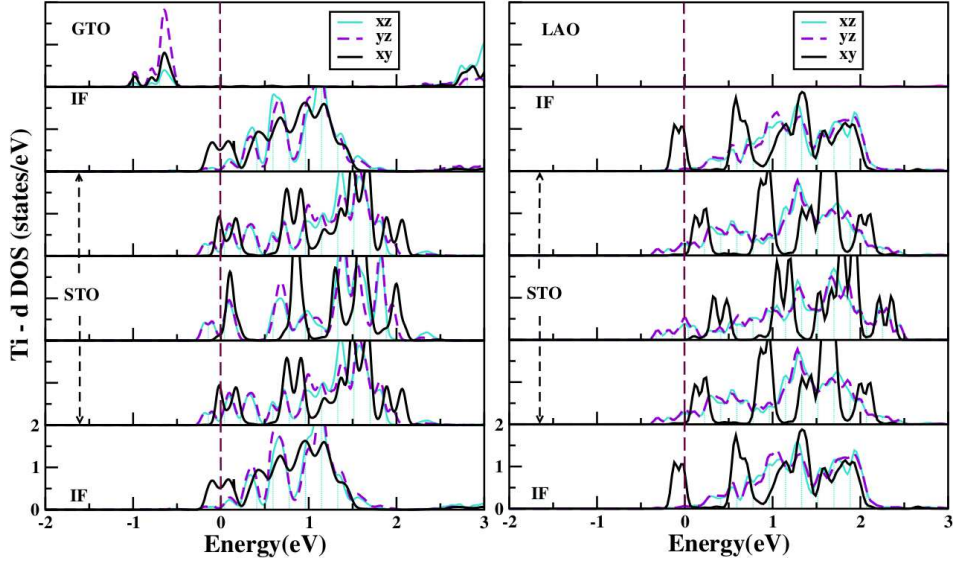}
\caption{The density of states projected to $xy$ (solid line), $xz$ (shaded area), $yz$  (dashed lines) orbitals of Ti, in different 
TiO$_2$ layers of (LAO)$_{1.5}$/(STO)$_{4.5}$ (right panels) and (GTO)$_{1.5}$/(STO)$_{4.5}$ (left panels) superlattices. 
The top panel corresponds to TiO$_2$ in GTO on the left, and to AlO$_2$ layer in LAO on the right
(for which there are no states in the energy range shown).
The subsequent panels from top to bottom (both on the left and right) refer to TiO$_2$ at 1st IF, 
1$^{st}$, 2$^{nd}$ and 3$^{rd}$ TiO$_2$ layers in STO block, and TiO$_2$ at the 2$^{nd}$ IF in the cell. 
The cell is symmetric about the 2nd TiO$_2$ layer in STO. The zero of the energy is the Fermi energy.}
\end{figure}

\begin{figure}
\includegraphics[width=0.9\columnwidth]{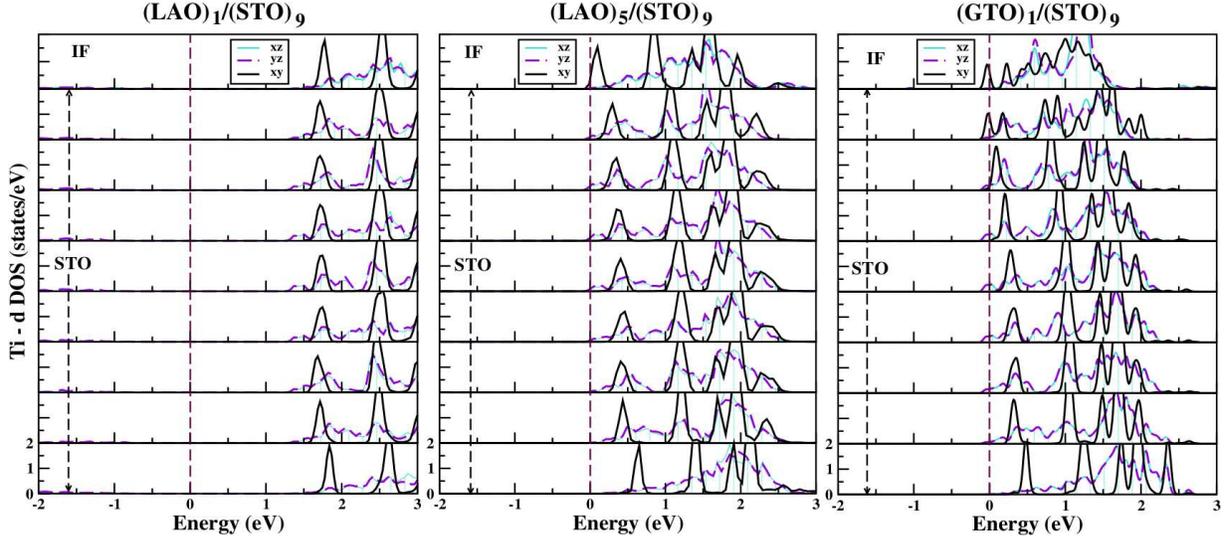}
\caption{The density of states projected to $xy$ (solid line), $xz$ (shaded area), $yz$  (dashed lines) orbitals of Ti, 
in different TiO$_2$ layers of LAO/STO and GTO/STO in thin film-substrate geometry.  
From top to bottom, different panels refer to TiO$_2$ at IF, TiO$_2$ layers 
belonging to STO block. The zero of the energy is set at E$_F$.}
\end{figure}

\begin{figure}
\includegraphics[width=0.85\columnwidth]{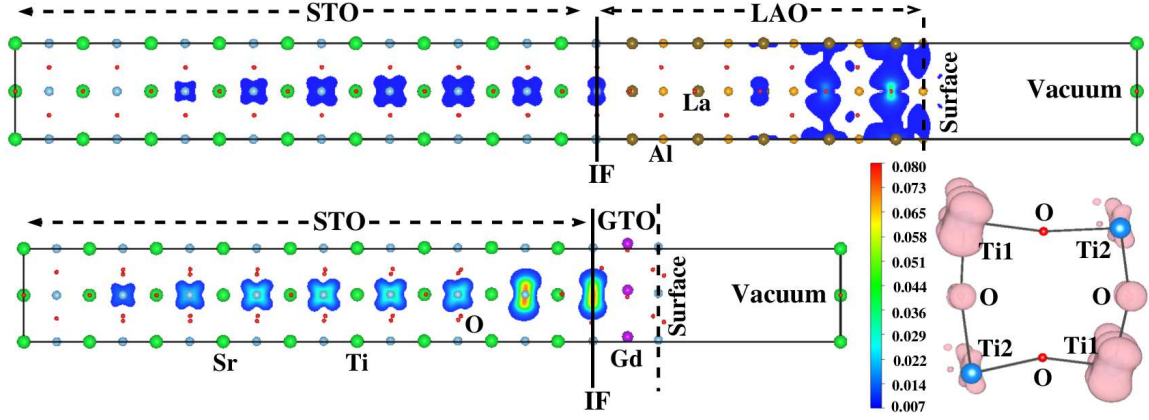}
\caption{Conduction electron charge density for (LAO)$_5$/(STO)$_9$ (upper panel) and 
(GTO)$_1$/(STO)$_9$ (lower panel) in thin film-substrate geometry, integrated over an energy window
from 0.1 eV below E$_F$ to E$_F$. The color of different contours corresponds to the values shown in the scale bar. 
The topmost AlO$_2$ surface layer in LAO is metallic, while the topmost TiO$_2$ layer in GTO is insulating.
Inset: We also show the charge disproportionation with two inequivalent Ti sites on the top layer of GTO
which drives the surface insulating. The charge density in this plot is obtained from integration over the 
lower Hubbard band in GTO; see text for details.}
\end{figure}

\begin{figure}
\includegraphics[width=0.85\columnwidth]{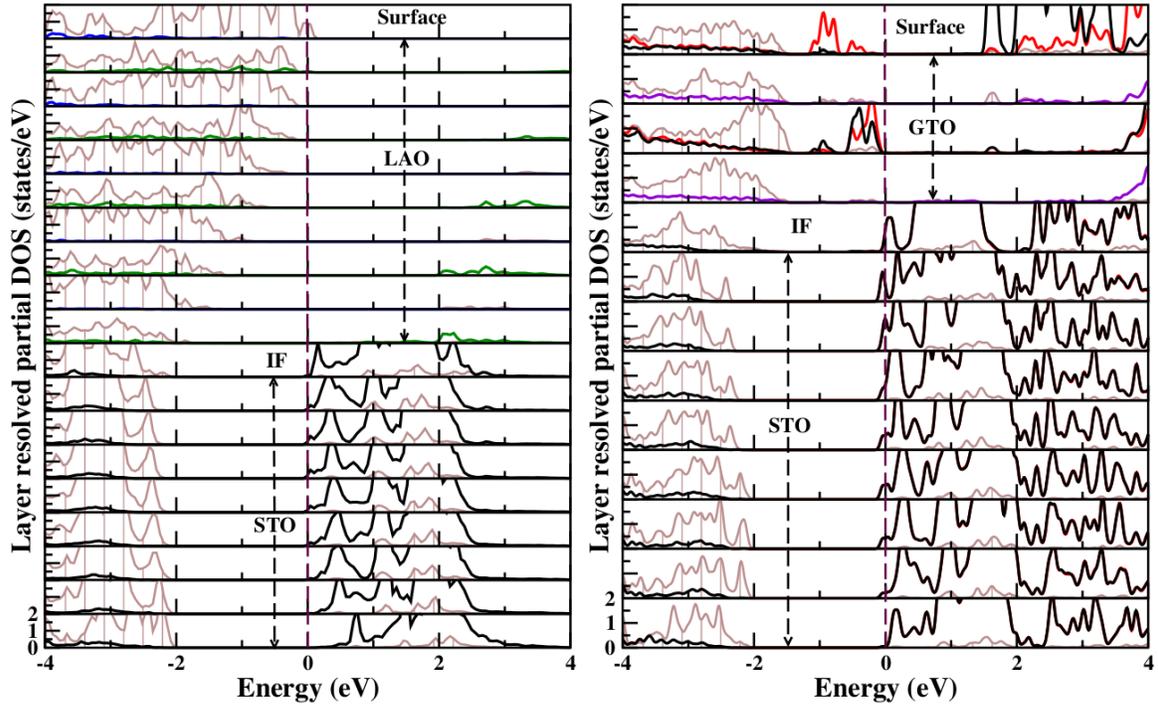}
\caption{The layer decomposed DFT density of states for (LAO)$_{5}$/(STO)$_9$ (left panel) and 
(GTO)$_{2}$/(STO)$_9$ (right panel) in thin film-substrate geometry, projected onto 
O $p$ (brown shaded), Al $p$ (green line), Ti $d$ (black/red line) and Gd $d$ (magenta) and
La $d$ (blue line) states. For GTO/STO, the projection to two charge disproportionated Ti atoms
are shown as black and red lines, respectively, while for LAO/STO projection to Ti $d$ is shown
as black line. E$_F$ is set at zero. From top to bottom, various panels show the surface layer, the
LAO or GTO block, the IF and the STO block.}
\end{figure}

\begin{figure}
\includegraphics[width=0.85\columnwidth]{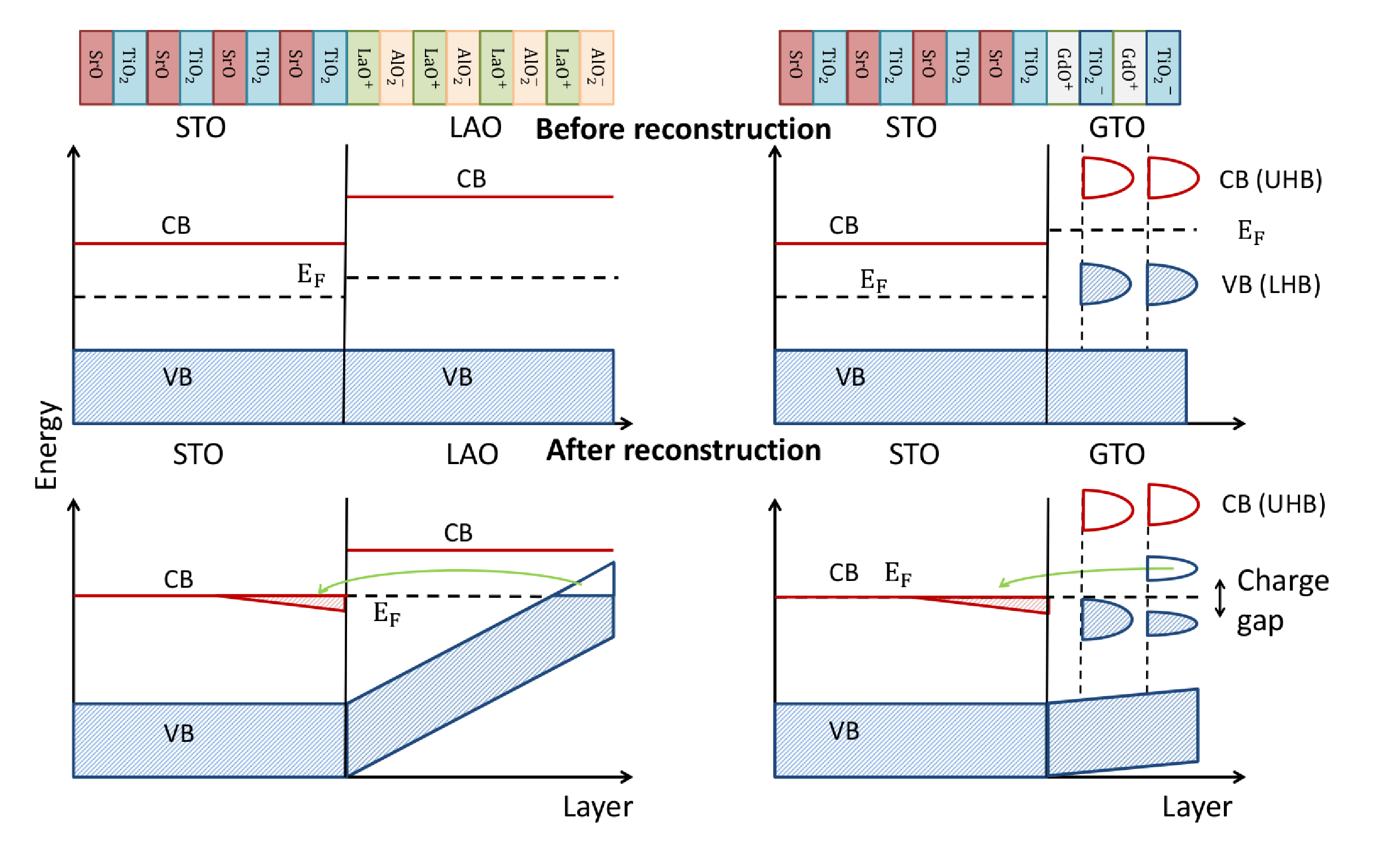} 
\caption{Schematic plot of the band offset and electronic reconstruction in LAO/STO (left panels) and GTO/STO (right panels). 
The upper edge VB of LAO and  STO derived from O $p$ bands are almost aligned, thereby causing a large band bending required 
for conduction. The upper edge VB of GTO and STO, on the other hand are completely misaligned, one being Ti $d$ lower Hubbard 
band (GTO), and another being 
O $p$ band (LAO). This makes band bending required for conduction much smaller as discussed in the text.}   
\end{figure}


\begin{thebibliography}{99}
\bibitem{hwang} Ohtomo, A. \& Hwang, H. Y. A high-mobility electron gas at the LaAlO$_3$/SrTiO$_3$ heterointerface. Nature (London) {\bf 427,} 423-426 (2004).
\bibitem{manhart} Mannhart, J. et al. Two-Dimension Electron Gases at Oxide Interfaces. MRS Bull, {\bf 33,} 1027-1034 (2008). 
\bibitem{huijben} Huijben, M. et al. Structure–Property Relation of SrTiO$_3$/LaAlO$_3$
Interfaces.  Adv. Mater. {\bf 21,} 1665-1667 (2009).
\bibitem{review1} Zubko, P. et al. Interface Physics in Complex Oxide Heterostructures.
Annu. Rev. Cond. Mat. Phys. {\bf 2,} 141-165 (2011).
\bibitem{review2} Sulpizio, A., Ilani, S., Irvin, P. \& Levy, J. Nanoscale Phenomena in Oxide Heterostructures. Ann. Rev. of Mater. Res. {\bf 44,} 117-149 (2014).
\bibitem{brinkman} Brinkman, A. et al, Magnetic effects at the interface between non-magnetic oxides. Nat. Mater. {\bf 6,} 493-496 (2007). 
\bibitem{reyren} Reyren, N. et al. Superconducting Interfaces Between Insulating Oxides. Science {\bf 317,} 1196-1199 (2007).
\bibitem{bert1} Bert, J. A. et al. Direct imaging of the coexistence of ferromagnetism and superconductivity at the LaAlO$_3$/SrTiO$_3$ interface. Nature Phys. {\bf 7,} 767-771 (2011).
\bibitem{li} Li, L., Richter, C., Manhart, J. \& Ashoori, R. C. Coexistence of magnetic order and two-dimensional superconductivity at LaAlO$_3$/SrTiO$_3$ interfaces. Nat. Phys. {\bf 7,} 762-766 (2011).
\bibitem{siemons} Siemons, W. et al. Origin of Charge Density at LaAlO$_3$ on SrTiO$_3$ Heterointerfaces: Possibility of Intrinsic Doping. Phys. Rev. Lett {\bf 98,} 196802 (2007).
\bibitem{basletic} Basletic, M. et al. Mapping the spatial distribution of charge carriers in LaAlO$_3$ /SrTiO$_3$ heterostructures. Nat. Mater. {\bf 7,} 621-625 (2008).
\bibitem{thiel} Thiel, S. et al. Tunable Quasi-Two-Dimensional Electron Gases in Oxide Heterostructures. Science {\bf 313,} 1942-1945 (2006).
\bibitem{moetakef} Moetakef, P. et al. Electrostatic carrier doping of GdTiO$_3$/SrTiO$_3$ interfaces. Appl. Phys. Lett. {\bf 99,} 232116 (2011).
\bibitem{cain} Cain, T. A., Moetakef, P., Jackson C. A. \& Stemmer, S. Modulation doping to control the high-density electron gas at a polar/non-polar oxide interface. Appl. Phys. Lett. {\bf 101,} 111604 (2012).
\bibitem{moetakef1} Moetakef, P. et al. Quantum oscillations from a two-dimensional electron gas at a Mott/band insulator interface. Appl. Phys. Lett. {\bf 101,} 151604 (2012).
\bibitem{moetakef2} Moetakef, P. et al. Carrier-Controlled Ferromagnetism in SrTiO$_3$. Phys. Rev. X {\bf 2,} 021014 (2012).
\bibitem{moetakef3} Moetakef, P. et al. Toward an artificial Mott insulator: Correlations in confined high-density electron liquids in SrTiO$_3$. Phys. Rev. B {\bf 86,} 201102 (R) (2012).
%
\bibitem{satpathy} Popovic, Z. S., Satpathy, S. \& Martin, R. M. Origin of the Two-Dimensional Electron Gas Carrier Density at the LaAlO$_3$ on SrTiO$_3$ Interface. Phys. Rev. Lett, {\bf 101,} 256801 (2008).
\bibitem{lao-elec} Zhong, Z. \& Kelly, P. J. Electronic-structure–induced reconstruction and magnetic ordering at the LaAlO$_3$|SrTiO$_3$ interface. Eur. Phys. Lett. {\bf 84,} 27001 (2008).
\bibitem{pentcheva} Pentcheva, R. \& Pickett, W. Avoiding the Polarization Catastrophe in LaAlO$_3$ Overlayers on SrTiO$_3$(001) through Polar Distortion. Phys. Rev. Lett. {\bf 102,} 107602 (2009).
\bibitem{chen} Chen, H., Kolpak, A. \& Ismail-Beigi, S. First-principles study of electronic reconstructions of LaAlO$_3$/SrTiO$_3$ heterointerfaces and their variants. Phys. Rev. B {\bf 82,} 085430 (2010).
%
\bibitem{chen2013} Chen, R., Lee, S. \& Balents, L. Dimer Mott insulator in an oxide heterostructure Phys. Rev. B {\bf 87,} 161119(R) (2013).
\bibitem{walle} Bjaalie, L., Janotti, A., Himmetoglu, B., \& Van de Walle, C. G. Turning SrTiO$_3$ into a Mott insulator. Phys. Rev. B {\bf 90,} 195117 (2014).
\bibitem{lechermann} Lechermann, F., \& Obermeyer, M. Towards Mott design by $\delta$-doping of strongly
correlated titanates. arXiv:1411.1637v2
\bibitem{janotti} Janotti, A., Bjaalie, L., Gordon, L. \& Van de Walle, C. G. Controlling the density of the two-dimensional electron gas at the SrTiO$_3$/LaAlO$_3$ interface. Phys. Rev. B {\bf 86,} 241108(R) (2012).
%
\bibitem{lit} Kim, J.S. et al. Nonlinear Hall effect and multichannel conduction in LaTiO3/SrTiO3 superlattices.
Phys. Rev. B {\bf 82,} 201407(R) (2010)
\bibitem{zhang01} Zhang, J. Y. et al. Correlation between metal-insulator transitions and structural distortions in high-electron-density SrTiO3 quantum wells. Phys. Rev. B {\bf 89,} 075140 (2014).
\bibitem{zhang02} Zhang, J. Y. Symmetry Lowering in Extreme-Electron-Density Perovskite Quantum Wells. Phys. Rev. Lett. {\bf 110,} 256401 (2013).
\bibitem{prl_2010} Dubroka, A. et al. Dynamical Response and Confinement of the Electrons at the LaAlO$_3$/SrTiO$_3$ Interface.  Phys. Rev. Lett. {\bf 104,} 156807 (2010). 
%
\bibitem{adv} Huijben, M. et al. Structureâ Property Relation of SrTiO$_3$/LaAlO$_3$ Interfaces. Adv. Mater. {\bf 21,} 1665-1677 (2009). 
\bibitem{zunger} L. Yu and A. Zunger, Nature Commun. {\bf 5}, 5118 (2014).
%
\bibitem{millis} 
Millis, A. J. Lectures on Oxide Interfaces at Ecole Polytechnique (2012), unpublished;
\\
http://phys.columbia.edu/$\sim$millis/oxideinterfaces/oxideInterfaces$\underline{\  }$1.html
%
%
\bibitem{kresse} Kresse, G. \& Hafner, J. Ab initio molecular dynamics for liquid metals. Phys. Rev. B {\bf 47,} 558(R) (1993).
\bibitem{kresse01}Kresse, G. \& J. Furthm$\ddot{u}$ller, J. Efficient iterative schemes for ab initio total-energy calculations using a plane-wave basis set. Phys. Rev. B {\bf 54,} 11169 (1996).
\bibitem{blochl} Bl$\ddot{o}$chl, P. E. Projector augmented-wave method. Phys. Rev. B {\bf 50,} 17953 (1994).
\bibitem{pbe} Perdew, J. P., Burke, K. \& Ernzerhof, M. Generalized Gradient Approximation Made Simple. Phys. Rev. Lett., {\bf 77,} 3865 (1996).
\bibitem{dudarev} Dudarev, S. L. el al. Electron-energy-loss spectra and the structural stability of nickel oxide: An LSDA+U study. Phys. Rev. B {\bf 57,} 1505 (1998).
%
\bibitem{imada}
Imada, M., Fujimori, A. \& Tokura, Y. Metal-insulator transitions. Rev. Mod. Phys. {\bf 70,} 1039 (1998).
\bibitem{valenti}
Zhang, Y., Jeschke., H. O. \& Valenti, R. Microscopic model for transitions from Mott to spin-Peierls insulator in TiOCl.
Phys. Rev. B {\bf 78,} 205104 (2008).

\end{thebibliography}
\end{document}